\documentclass{statsoc}

\usepackage[a4paper]{geometry}
\usepackage{natbib}
\usepackage{hyperref}
\usepackage{amsmath}

\usepackage{upgreek}

\usepackage{tikz}

\usetikzlibrary{decorations.pathmorphing} 
\usetikzlibrary{fit}		
\usetikzlibrary{backgrounds}	
\usetikzlibrary{positioning}
\usepackage{bm}

\usepackage{multirow}

\title[Missing data analysis with latent GMRFs]{Missing data analysis and imputation via latent Gaussian Markov random fields}

\author[G\'omez-Rubio {\it et al.}]{Virgilio G\'omez-Rubio}
\coaddress{Department of Mathematics,
School of Industrial Engineering,
University of Castilla-La Mancha,
02071 Albacete, Spain.
}
\email{Virgilio.Gomez@uclm.es}

\author[G\'omez-Rubio {\it et al.}]{Michela Cameletti}
\address{Dept. of Management, Economics and Quantitative Methods,
University of Bergamo,
Bergamo, IT}

\author[G\'omez-Rubio {\it et al.}]{Marta Blangiardo}
\address{Dept. of Epidemiology and Biostatistics, Imperial College London,
London, UK}

\begin{document}

\maketitle

\begin{abstract}

In this paper we recast the problem of missing values in the covariates of a regression model as a
latent Gaussian Markov random field (GMRF) model in a fully Bayesian framework.  Our
proposed approach is based on the definition of the covariate imputation sub-model
as a latent effect with a GMRF structure.  We show how this formulation works
for continuous covariates and provide some insight on how this could be extended
to categorical covariates.

The resulting Bayesian hierarchical model naturally fits within the integrated
nested Laplace approximation (INLA) framework, which we use for
model fitting.  Hence, our work fills an important gap in the INLA methodology
as it allows to treat models with missing values in the covariates.

As in any other fully Bayesian framework, by relying on INLA for model fitting it is possible to formulate a
joint model for the data, the imputed covariates and their missingness
mechanism. In this way, we are able to tackle the more general problem of
assessing the missingness mechanism by conducting a sensitivity analysis on the
different alternatives to model the non-observed covariates.

Finally, we illustrate the proposed approach with two examples on modeling
health risk factors and disease mapping. Here, we rely on two different
imputation mechanisms based on a typical multiple linear regression and a
spatial model, respectively. Given the speed of model fitting with
INLA we are able to fit joint models in a short time, and to easily conduct
sensitivity analyses.

\end{abstract}


\section{Introduction}

In any statistical analysis missing data is one of the most important issues a
researcher needs to deal with; failing to properly account for it can result in
a reduction of statistical power, or even in biased statistical inference.
Consequently, countless methods have focused on this \citep[see, for
example,][]{Schafer:1997, LittleRubin:2002, Enders:2010, Buuren:2012,
Raghunathan:2015}. 

Missing data can occur for a number of reasons, as described in
\cite{LittleRubin:2002}. Sometimes, the missingness mechanism is ignorable and
inference can rely on the observed data alone, appropriately coupled with a suitable
imputation or data augmentation model if needed. When the missingness mechanism
is not ignorable, a joint approach is required to fit the analysis model,
impute the missing values and assess the missingness mechanism. Under
this scenario, it is recommended that a sensitivity analysis is carried out to assess the
impact of the missingness mechanism on the model parameters estimates \citep{mason_strategy_2012}.

The Bayesian paradigm has gained popularity for dealing with missing data,
making no distinction between parameters and missing data which are considered
as additional unknown parameter with a prior distribution. For these reasons
and differently from other ad-hoc methods \citep{Nakagawa:2015}, with a full
Bayesian approach it is possible to combine together the analysis and
imputation model in a joint estimation framework \citep{Erler:2016}.  For
instance, \cite{mason_bayesian_2009} and \cite{mason_strategy_2012} developed a
fully Bayesian missing imputation framework in order to adjust for
several missing covariates in longitudinal or cross-sectional studies; each of
the missing covariates is assigned an imputation model, all 
jointly modelled with the analysis model.

The approach we propose in this paper is based on recasting the imputation
sub-model to define it as a latent Gaussian Markov random field
\citep[GMRF,][]{GMRF:2005}  which is part of a larger Bayesian hierarchical
model. This fits naturally within the integrated nested Laplace approximation
\citep[INLA,][]{Rueetal:2009} methodology,  as an alternative to Markov chain
Monte Carlo (MCMC). This approach is suitable for continuous covariates and can
be also extended to impute categorical variables.
 This makes model fitting with
missing covariates possible in INLA, and our new approach fills an important
gap, as INLA has always required the data in the latent GMRF that
defines the model to be fully observed.  Here we focus on missing covariates only as
INLA  can
easily fit models with missing data in the response variable, simply 
computing the corresponding posterior predictive distribution derived from the
analysis model to be fit \citep{VirBook:2020}. 

A previous attempt to solve the issue of missing values in the covariates in
the INLA framework can be found in \cite{VirRue:2018}. They adopt a Gaussian
prior for the imputation of the missing values in the covariates and sample
from the  missing data posterior distribution through INLA within MCMC. 
A different approach is proposed in Chapter 8 of \cite{INLABook:2015}, where a
bivariate model for spatially misaligned data is estimated by adopting the
stochastic partial differential equations (SPDE) approach
\citep{Lindgren:2011}. Covariate values are imputed (in new locations) by
assuming a spatial Gaussian field which is also included in the linear
predictor of the response model.  Alternatively \cite{VirBook:2020} proposes a multiple
imputation (MI) approach \citep{rubin_multiple_1987,rubin_multiple_1996,CarpenterKenward:2012} so
that the covariates are imputed multiple times through resampling so that $N$ complete
datasets are used in the analysis model. All the results are then combined 
to
obtain the final estimates of the model parameters 
\citep[see][for details]{rubin_multiple_1987}.
We differ from the previous approaches in that we propose a joint framework, similarly to \cite{mason_strategy_2012}. Through the joint model, the uncertainty about the imputation of the missing covariates
propagates throughout the model so that it also reflects on the model
parameter estimates in the analysis. At the same time, information from the outcome in the analysis model feedback on the imputation, making it un-necessary to include the outcome in the imputation model, as commonly done in the classic MI approach. Our new approach fits naturally within the
INLA framework, can be extended to consider different types of problems (i.e.,
not only spatial models) and can be easily fit with the associated R-INLA
package for the R programming language \citep{VirBook:2020}.

The paper is structured as follows. In Section~\ref{sec:missingvalues} we
review methods for missing values, while in Section~\ref{sec:GMRF_imputation}
we introduce a novel method for missing values imputation.
 Section~\ref{sec:INLA} presents a
brief summary of the INLA approach to Bayesian inference and how our novel
approach fits within this framework.  Section~\ref{sec:examples} shows a two
examples for the application of our proposed method and
Section~\ref{sec:discussion} presents discussion points.

%

\section{Approaches to deal with missing data}
\label{sec:missingvalues}

In their seminal book, \cite{LittleRubin:2002} identify three possible
mechanisms of missingness. If the probability of being missing is the same for
all the observations, we can assume that the missing data distribution does not
depend on any of the observed or missing variables.  In this case the data are
said to be  \textit{missing completely at random} (MCAR). If the distribution
of the missing data depends on completely observed variables (and do not depend
on the missing ones), the data are called \textit{missing at random} (MAR). An
example  of MAR is that women are less likely to answer questions related to
their income than men, but this has nothing to do with the income itself.
Finally, if neither MCAR or MAR holds the \textit{missing not at random} (MNAR)
case occurs and the missing values distribution depends on both missing and
observed variables. For instance, in a neurological questionnaire, a subject is less likely to answer questions related to the disease if this is severe.

Under MCAR or MAR, 
the missing data mechanism is \textit{ignorable}.
As reported in \cite{Seaman:2013} this means that inferences obtained from a
parametric model for the observed data alone are the same as inferences
obtained from a joint model for the data and missingness mechanism.  On the
contrary, if the data are MNAR the missing data mechanism is not ignorable and
a model for the missingness mechanism is required.  It is important to note that we cannot gather evidence from
the data at hand about the missing data mechanism (MCAR, MNAR or MAR). On the
basis of the knowledge regarding the data collection methods and the assumed relationship among the collected variables, it is
possible only to make assumptions about the reasons for missing data, choose
the best corresponding strategy for data analysis \citep{Pigott:2001} and 
conduct a sensitivity analysis on these assumptions
\citep{mason_strategy_2012}.

The simplest and most popular ad-hoc methods to deal with missing information
consists in replacing the missing data with a plausible value, such as the mean
or median calculated over the observed cases (or the mode if the variable is
categorical) or to perform a complete cases analysis (i.e., removing the
observations with one or more missing values). However, while the first method
has the potential of distorting the data distribution and to underestimating
their variability, the second one has the major drawback of reducing the power
of the study (as the dataset for the analysis will have a reduced size) and of
producing biased estimates if the MCAR assumption is not valid. To overcome
this issue, inverse probability weighting was developed, based on the idea of
assigning different weights to the different complete cases based on specific
characteristics which are relevant for the missing data; in  two reviews
\cite{Carpenteretal:2006,Seaman:2013} showed advantages and drawbacks of such approach.

In the last three decades model-based methods have been preferred to account
for missing data in the case of an ignorable missing data mechanism; see for
instance the papers by \citealt{little_regression_1992};
\citealt{little_statistical_1987}; \citealt{schafer_missing_2002}. Regression
mean imputation is the simplest of the model-based methods, where the variable
with missing data is predicted based on a regression model which includes the
other variables as regressors. To overcome the issue of unreasonably lack of
uncertainty for the imputed values, stochastic regression imputation was
proposed, to generate imputed values adding some random noise
\citep{Nakagawa:2015}.

A well established and increasingly popular model-based approach to dealing with missing data occurring in more than one variable is multiple imputation (MI) proposed by \cite{rubin_multiple_1987,rubin_multiple_1996}. Through Monte Carlo simulation, it produces several versions of the complete dataset which differ by the imputed missing values. Then, for each complete dataset the estimates of interest are computed (by fitting a \textit{substantive} model) and the results are pooled together into a final estimate which takes into account the uncertainty of the imputed data. The imputation of the missing values can be done using mainly two strategies \citep{Buuren:2012}: i) \textit{joint modeling}, when missing values are imputed by sampling from a multivariate model fitted to the data (usually the multivariate Gaussian is used); ii) \textit{fully conditional specification} \citep[also known as multiple imputation using chained equation, MICE,][]{micepkg}, when conditional univariate distributions are used to impute the missing values iteratively through a variable-by-variable approach (see \citealt{white_multiple_2011} for a thorough review of this method).

\subsection{Bayesian inference}

Bayesian inference provides a suitable framework for dealing with missing data,
as it treats missing data similarly to model parameters, making no distinction
between them.  For these reasons and differently from other methods, with a
full Bayesian approach, it is possible to include the main model for data
analysis, imputation of missing values and a missingness model in a joint
estimation framework \citep{Erler:2016}.

Let $\bm{z}$ denote the \textit{complete} set of data. We assume that $\bm{z}
=(\bm{z}_{obs},\bm{z}_{mis})$, where $\bm{z}_{obs}$ denote the observed  values
while $\bm{z}_{mis}$ refer to the missing values. Moreover, let $\bm{M}$ be the
missing data indicator variable, i.e. a vector or matrix with the same length
or dimension as $\bm{z}$ with values equal to 1 (or 0) if the corresponding
values of $\bm{z}$ is missing.  

Following the selection model approach \citep{Nakagawa:2015} the joint
distribution of $\bm{z}$, $\bm{M}$, the model parameters $\bm\theta_z$ and the parameters
in the missingness model $\bm{\theta}_M$ can be expressed as

$$
\pi(\bm{z}, \bm{M}, \bm{\theta}_z, \bm{\theta}_M) = \pi(\bm{z}, \bm{\theta}_z) \pi(\bm{M} \mid \bm{z}, \bm{\theta}_M).
$$ 
\noindent
This formulation assumes that parameters $\bm\theta_z$ and $\bm\theta_M$ are
distinct and with independent priors. 

Following this, $\pi(\bm{M} \mid \bm{z}_{obs}, \bm{z}_{mis},\bm\theta_M)$ depends on a set of parameters $\bm\theta_M$, and models the missing data mechanism for the three cases introduced above \citep{LittleRubin:2002}:
\begin{description}
\item[MCAR] if the distribution does not depend on any of the fully or partiallyobserved variables, i.e. $\pi(\bm {M}\mid \bm{z}_{obs}, \bm{z}_{mis},\bm\theta_M)=\pi(\bm{M} \mid \bm\theta_M)$.  
\item[MAR] if the distribution depends only on fully observed variables, which
means that $\pi(\bm{M}\mid \bm{z}_{obs}, \bm{z}_{mis},\bm\theta_M)=\pi(\bm{M}\mid \bm{z}_{obs}, \bm\theta_M)$. This implies that, given the observed data, the missingness mechanism does not depend on the unobserved data.
\item[MNAR] if the distribution $\pi(\bm{M}\mid \bm{z}_{obs}, \bm{z}_{mis},\bm\theta_M)$ depends on fully and partially observed variables.
\end{description}

If the data are MCAR or MAR and the parameters $\bm\theta_M$ are distinct of the
parameters of the data generating process, $\bm\theta_z$, and with independent
priors, then the missing data
mechanism is \textit{ignorable} and and $\pi(\bm{M}\mid \bm{z}_{obs}, \bm{z}_{mis},\bm\theta_M)$ can
be omitted \citep{Seaman:2013}. On the contrary if the data are MNAR the missing data mechanism is not ignorable and a model for missingness is required (i.e. a logistic model)
and has to be jointly estimated with the main model, that will
include an imputation sub-model for the missing values.

Note that we can not tell from the data at hand whether the missing observations are MCAR, MNAR or MAR and at the same time it is not trivial to specify a model of missingness.  In this case, a sensitivity analysis needs to be carried out to assess the impact of different scenarios for the missing data on the estimates of the model parameters \citep{Carpenteretal:2007,mason_strategy_2012}.

%
%
%
%

\subsection{Missing data in the response variable}

Let $\bm{z}=(\bm{y},\bm{x})$ be the set of data including the response $\bm{y}$
and the covariates $\bm{x}$. If we assume that the covariates are fully
observed, we have that $\bm{z}_{obs} =(\bm{y}_{obs},\bm{x})$ and $\bm{z}_{mis}
=(\bm{y}_{mis})$. In this case the likelihood
$\pi(\bm{z}_{obs},\bm{z}_{mis}\mid \bm\theta_z)$ corresponds to the
distribution of $\pi(\bm{y}_{obs},\bm{y}_{mis}\mid \bm{x}, \bm\theta_y)$, with
$\bm\theta_y$ the hyperparameters in the likelihood.

If we assume that the missing data mechanism is ignorable, the imputation of
the missing data values $\bm{y}_{mis}$ is simply done through the posterior
predictive distribution $p(\bm{y}_{mis}\mid \bm{y})$.  In general, we will have the
observation model by defining an appropriate distribution for the likelihood.
In addition, the mean of observation $i$, $\phi_{i}$ ,will be linked to
a linear predictor on the covariates and other effects using an appropriate
link function $g(\cdot)$, i.e.,

\begin{equation}
\label{Eq:obsmodel} g(\phi_{i})= \beta_0 +
\sum_{p=1}^P \beta_p x_{pi} + \sum_{l=1}^L f_l(z_{li}) 
\end{equation}
\noindent
Here, $\beta_0$ is an intercept, $\{\beta_p\}_{p=1}^P$ the coefficients of the
$P$ covariates available $\{\bm{x}_p\}_{p=1}^P$ and $\{f_l(\cdot)\}_{l=1}^L$ represent
$L$ different non linear effects on covariates $\{\bm{z}_l\}_{l=1}^L$.


If instead data are MNAR a missing mechanism model $\pi(\bm{M}\mid \bm{y},\bm{x},\bm\theta_M)$ is
required, e.g.
\begin{eqnarray}\label{Eq:missingmodel}
M_i  &\sim& Bernoulli(p_i)\nonumber\\ 
logit(p_i) &=& \gamma_0 + \sum_{r=1}^R \gamma_r x_{ri} + \delta y_{i}
\end{eqnarray}
where $\bm\theta_M=(\gamma_1, \gamma_1,\ldots,\gamma_R, \delta)$ and $M_i$
is a missingness indicator for $y_i$. In addition,
an imputation model for the missing values will be required.

%
%

However, in this work we will assume that there are no missing observations in
the response or that the missingness mechanism is ignorable, which means that
posterior inference is based on the predictive distribution.

\subsection{Missing data in the covariates}

We now consider the case when $\bm{z}_{obs} =(\bm{y},\bm{x}_{obs})$ and $\bm{z}_{mis} =(\bm{x}_{mis})$,
with $\bm{x}_{obs}$ the observed values of the covariates and $\bm{x}_{mis}$ the missing
ones. Distribution $\pi(\bm{y}, \bm{z}_{obs},\bm{z}_{mis}\mid \bm\theta_z)$ can be written as

\[
\pi(\bm{y},\bm{x}_{obs},\bm{x}_{mis}\mid \bm\theta_z)=\pi(\bm{y}\mid \bm{x}_{obs},\bm{x}_{mis},\bm\theta_y)\pi(\bm{x}_{obs},\bm{x}_{mis}\mid \bm\theta_x)
\]
assuming that $\bm\theta_z=(\bm\theta_y,\bm\theta_x)$ is the vector of conditionally
independent parameters. The distribution $\pi(\bm{x}_{obs},\bm{x}_{mis}\mid \bm\theta_x)$
represents the joint distribution of observed and missing covariates and it
includes the imputation model. For example, the joint distribution can be a
multivariate normal distribution (taking into consideration correlation between
covariates) for continuous covariates, or a discrete distribution if the
covariate is categorical.

In general, we will have the observation model as in
equation~\eqref{Eq:obsmodel} together with the imputation model and the
missingness model (described in Section \ref{sec:GMRF_imputation}) as in equation~\textcolor{red}{~\eqref{Eq:missingmodel}} but only if the
missing data are MNAR. 


%

%

\section{Imputation of continuous missing covariates}
\label{sec:GMRF_imputation}


For simplicity, we will consider the imputation of a single covariate with
missing observations, but this approach can be easily extended to consider the
imputation of missing values in several continuous covariates using a
multivariate model.

We define a latent effect
$\bm{x}^{\prime}$ as a GMRF with mean $\bm\mu^{\prime}(\bm\theta_I)$
and precision $\bm{Q}^{\prime}(\bm \theta_I)$, with $\bm\theta_I$ being the set
of hyperparameters related to the imputation procedure.  The latent effect is
split in two parts $\bm{x}^{\prime} = (\bm{x}^{\prime}_{mis},
\bm{x}^{\prime}_{obs})$. The distribution of  $\bm{x}^{\prime}_{obs}$ is
assumed to be as close as possible to the observed covariate data
$\bm{z}_{obs}$; to guarantee this, we set the mean equal to $\bm{z}_{obs}$ and
a high precision (e.g., $10^{10}$). An imputation model is built for 
$\bm{x}^{\prime}_{mis}$ (with mean $\bm\mu_c$ and precision 
$\bm Q_c$), depending on $\bm \theta_I$ and whose details will be given in
Section \ref{Sect:genericimputationmodel}. Finally, we will also assume that
$\bm{x}_{obs}^{\prime}$ and $\bm{x}_{mis}^{\prime}$ are independent given the
latent effect hyperparameter $\bm\theta_I$.  Consequently the joint
distribution of $\bm x^\prime$ is given by

\begin{equation}\label{Eq:newlatenteff}
\bm x^\prime \mid \bm \theta_I
\sim \text{Normal}
\Bigg(
\left(\begin{array}{c} \bm \mu_{c} \\ \bm z_{obs}\end{array}\right),
\left[\begin{array}{cc} \bm Q_c & \bm 0\\ \bm 0 &10^{10} \bm I
\end{array}\right]
\Bigg).
\end{equation}



\subsection{Imputation model}\label{Sect:genericimputationmodel}


Differently from Section~\ref{sec:missingvalues}, let $\bm z=(\bm z_{mis},\bm z_{obs})$ denote the complete set of values of 
the covariate and we will write the response values $\bm{y}y$ separately where needed.
The \textit{imputation model} will provide the distribution of the missing
values $\bm{z}_{mis}$ given the observed data $\bm{z}_{obs}$ and the
hyperparameters $\bm\theta_I$ (that take values in the parameter space
$\Theta_I$). In a Bayesian framework, a 
sub-model is specified, where $\bm{z}_{obs}$ can be regarded as the data while $\bm{z}_{mis}$ and $\bm\theta_I$ the parameters to
estimate. Hence, we have

$$
\pi(\bm{z}_{mis} \mid \bm{z}_{obs}) =
\int_{\Theta_I} \pi(\bm{z}_{mis}, \bm\theta_I \mid \bm{z}_{obs}) d\bm\theta_I = 
\int_{\Theta_I} \pi(\bm{z}_{mis} \mid \bm{z}_{obs}, \bm\theta_I)
  \pi(\bm\theta_I \mid \bm{z}_{obs}) d\bm\theta_I.
$$

Here, $\pi(\bm{z}_{mis} \mid \bm{z}_{obs}, \bm\theta_I)$ is the
conditional distribution of the missing values given the observed data and the
hyperparameters of the imputation model. Also, $\pi(\bm\theta_I\mid 
\bm{z}_{obs})$ can be regarded as the posterior distribution of the
hyperparameters in the imputation sub-model given the observed data. Note
that this distribution is estimated only from the observed data $\bm{z}_{obs}$,
so it can be regarded as an \textit{informative prior} for $\bm\theta_I$.
Moreover,  it can be rewritten as
\[
 \pi(\bm\theta_I \mid \bm{z}_{obs}) \propto \pi(\bm z_{obs}\mid \bm\theta_I)\pi(\bm \theta_I)
\]
where $\pi(\bm{z}_{obs} \mid \bm\theta_I)$  is obtained by integrating
$\bm{z}_{mis}$ out in the distribution of $\bm{z}$. 
Finally, the hyperparameters $\bm\theta_I$  are typically modeled as exchangeable a priori.

As stated above, to derive the distribution for the imputation model $\pi(\bm{z}_{mis} \mid \bm{z}_{obs}, \bm\theta_I)$, we first assume a multivariate Normal distribution for the joint distribution of the complete set of covariates $\bm z$: 
\begin{equation}\label{Eq:jointz}
\bm z \mid \bm \theta_I
\sim \text{Normal}
\Bigg(
\left(\begin{array}{c} \bm \mu_{mis} \\ \bm \mu_{obs}\end{array}\right),
\left[\begin{array}{cc} \bm Q_{mis,mis} & \bm Q_{mis,obs} \\ \bm Q_{obs,mis} & \bm Q_{obs,obs}
\end{array}\right]
\Bigg)=\text{Normal}\left(\bm \mu,\bm Q\right),
\end{equation}
where both the mean and the precision matrix can depend on $\bm\theta_I$. It follows that the \textit{imputation model} is defined by the following conditional distribution \citep{GMRF:2005}:
\[
\bm z_{mis} \mid \bm z_{obs}, \bm\theta_I \sim \text{Normal}\left(\bm \mu_c, \bm Q_c\right)
\]
where $\bm\mu_c=\bm \mu_{mis} -\bm Q_{mis,mis}^{-1}\bm Q_{mis,obs} \left(\bm z_{obs}-\bm \mu_{obs}\right)$ and $\bm Q_c=\bm Q_{mis,mis}$. Note that $\bm \mu_c$ and $\bm Q_c$ are necessary to define the distribution of the new latent effect defined by equation~\eqref{Eq:newlatenteff}.

We now describe two particular examples of imputation with a
typical linear regression and spatial models (useful when the covariate is
spatially correlated). However, the principles presented below can be extended to a wide range
of models, including longitudinal data, time series and other smooth
terms.


\subsection{Imputation with a linear regression model}
\label{Sec:imputationlinreg}

The first imputation model that we describe is based on the linear
regression model. We assume that the mean of the multivariate Normal
distribution in equation~\eqref{Eq:jointz} is defined, considering the $n$
observations, as $\bm X \bm \beta^{\top}$.  Here, $\bm{X}$ is a matrix
of $p$ fully observed covariates (columnwise) with associated coefficient
vector $\bm \beta=(\beta_0, \ldots, \beta_{P})$. To match the structure of
$\bm z=(\bm z_{mis},\bm z_{obs})$, matrix $\bm{X}$ can be rewritten as a block
matrix as 

$$
\bm{X} = 
\left[
\begin{array}{c}
\bm{X}_{mis}\\
\bm{X}_{obs}\\
\end{array}
\right]
$$

Under the linear regression model, we assume that the mean of $\bm z$ depends on a linear
combination of the fully observed covariates, i.e. $\bm \mu=E(\bm z)=\bm{X}\bm \beta^{\top}$. By adopting the block notation, we thus assume the 
following joint distribution: 
\[
\bm z \mid \bm \theta_I
\sim \text{Normal}
\Bigg(
\left(\begin{array}{c} \bm X_{mis}\bm\beta^{\top} \\ \bm X_{obs}\bm \beta^{\top}\end{array}\right),
\left[\begin{array}{cc}\tau \bm I_{mis} & \bm 0 \\ \bm 0 &  \tau \bm I_{obs}
\end{array}\right]
\Bigg)
\]
where $\tau$ is the precision hyperparameter and $\bm{I}_{mis}$ and $\bm{I}_{obs}$ are identity matrices whose dimensions depend on the number of missing and observed data in $\bm z$. In this case the vector of hyperparameters
is given by $\bm\theta_I = (\bm\beta, \tau)$. Note that, given $\bm \theta_I$, observations
are assumed independent of each other, which simplifies the model.

Following the approach presented in Section \ref{Sect:genericimputationmodel}, we obtain that  the conditional distribution of $\bm{z}_{mis} \mid  \bm{z}_{obs}, \bm \theta_I $ (i.e. the imputation model) has
the following mean and precision:

$$
\bm\mu_c = \bm{X}_{mis}\bm\beta^{\top} \qquad \qquad \bm{Q}_c = \tau \bm{I}_{mis}
$$

As stated above, note that $\bm \beta$ and $\tau$ are informed by $\pi(\bm
\beta, \tau \mid \bm z_{obs})$, which is proportional to $\pi(\bm z_{obs} \mid \beta,
\tau) \pi(\beta, \tau)$.

Finally, priors must be set on the hyperparameters. For simplicity, each of the
elements in $\bm \beta$ is assigned a Normal distribution with zero mean and
large precision.  Parameter $\tau$ has a vague prior (e.g., a Gamma distribution
with large variance). 
All hyperparameters are independent a priori, so that
$\pi(\bm\theta_I) = \pi(\tau)\Pi_{i=0}^{P} \pi(\beta_i)$.  Note that other
priors could be easily considered here.


\subsection{Imputation with a spatial model}
\label{Sec:imputationCAR}

When the covariate to be imputed is spatially correlated
we can assume a conditional
autoregressive specification \cite[Chapter 13]{HBspatialstats:2010} so that
the mean is $\bm \mu = \bm\alpha^{\top} = (\alpha,\ldots,\alpha)^{\top}$  and
the precision is  $\bm
Q= \tau(\bm{I} - \rho \bm{W})$. Here, $\alpha$ is the intercept
of the linear predictor, $\rho$ is a spatial autocorrelation parameter, and
$\bm{W}$ is an adjacency matrix, defining the sets of neighbours. This is
often scaled
dividing it by its largest eigenvalue as this will allow us to take $\rho$  in
the $(0, 1)$ interval.
Note that $\bm{W}$ can be rewritten as a block matrix
with four sub-matrices according to missing and observed values, as  done with
$\bm{Q}$ in equation~\eqref{Eq:jointz}.  The vector of hyperparameters is now given
by $\bm\theta_I = (\tau, \rho, \alpha)$.

Adopting the block notation, under the CAR specification for imputation we thus
assume the following joint distribution for  $\bm z=(\bm z_{mis},\bm z_{obs})$: 

\[
\bm z \mid \bm \theta_I
\sim \text{Normal}
\Bigg(
\left(\begin{array}{c} \bm \alpha_{mis}^{\top} \\ \bm \alpha_{obs}^{\top}\end{array}\right),
\left[\begin{array}{cc} \tau(\bm{I}_{mis} - \rho \bm{W}_{mis,mis}) & \bm  - \tau\rho \bm{W}_{mis,obs} \\ \bm - \tau \rho \bm{W}_{obs,mis} & \tau(\bm{I}_{obs} - \rho \bm{W}_{obs,obs})
\end{array}\right]
\Bigg).
\]

It then follows that the conditional distribution of $\bm{z}_{mis} \mid
\bm{z}_{obs}, \bm \theta_I $ (i.e. the imputation model) is characterized by
the following mean and precision matrix:

$$
\bm\mu_{c} = \bm\alpha_{mis}^{\top}  - (\bm{I}_{mis} - \rho \bm{W}_{mis,mis})^{-1}
(- \rho \bm{W}_{mis,obs}) (\bm{z}^{\top}_{obs} - \bm\alpha_{obs}^{\top})
$$

$$
\bm{Q}_{c} = \tau\left(\bm{I}_{mis} - \rho \bm{W}_{mis,mis}\right)
$$

Again, $\tau$, $\rho$ and $\alpha$ are informed by $\pi(\tau, \rho, \alpha \mid \bm z_{obs})$, which is proportional to the product $\pi(\bm z_{obs} \mid \tau, \rho, \alpha) \pi(\tau, \rho, \alpha)$.

Finally, $\alpha$ is given a Gaussian prior with zero mean and small precision,
$\tau$ is assigned a vague prior (e.g., a Gamma distribution with a small precision), while
$logit(\rho)$ is
assigned a Gaussian prior with zero mean and small precision \citep[see, for
example,][Chapter 5, for details on why this parameterization is used]{VirBook:2020}.


\subsection{Extension to the imputation of categorical missing covariates}\label{sec:GMRF_imputation_categ}

The imputation of the missing values in categorical variables does not fit into
the GMRF framework described in Section \ref{sec:GMRF_imputation} as these
variables are defined in a discrete space. For this reason, a different
approach will be considered for defining the imputation model
$\pi(\bm{z}_{mis} \mid \bm{z}_{obs})$ and for estimating the model. In particular,
as imputation model we will consider a multinomial likelihood which can be
fit with INLA by using the multinomial-Poisson transformation
\citep{Baker:1994}. 

Note that in this case the procedure is similar to the multiple imputation
approach: the imputation model is specified where the categorical variables
with missing values are considered as the response variables, so that the
predictive distribution of the missing observations can be computed. Similarly
to the case of missing data in the response, values are sampled to fill the
missing values in the covariates. Then, the analysis model is run by using the
imputed covariates as completely known. This procedure is repeated by
simulating several samples and estimating the corresponding models; finally,
all the resulting models are averaged by using Bayesian model averaging \citep{VirRue:2018}. Note
that this approach does not produce feedback in the estimation of the
parameters of the imputation model as in the previous approach, given that it is done in two-stages rather than jointly. For this reason, and similarly to the classical MI, the outcome $\bm{y}$ should be included in the imputation model.  Alternatively,
INLA within MCMC can be used to fit the joint model using a fully Bayesian
approach \citep[see the example in ][]{VirRue:2018}.

Inference on the model parameters when multiple imputation of a categorical
covariate can be summarized as follows.  Considering the generic
parameter $\theta_k$ we can write its posterior marginal
distribution as: 

$$
\pi(\theta_k \mid \bm{z}_{obs}, \bm{y}) =
\sum_{\bm{z}_{mis}\in \Theta_{mis}} \pi(\theta_k, \bm{z}_{mis} \mid \bm{z}_{obs}, \bm{y}) =
\sum_{\bm{z}_{mis}\in \Theta_{mis}} \pi(\theta_k\mid \bm{z}_{obs},  \bm{z}_{mis}, \bm{y}) \pi(\bm{z}_{mis} \mid \bm{z}_{obs}, \bm{y}).
$$
\noindent
Here, $\Theta_{mis}$ represents the parametric space of the missing values
of the categorical covarite, which in a Bayesian framework are considered
to be random variables.

Given $L$ samples $\{\bm{z}_{mis}^{(l)}\}_{l=1}^L$ from 
$\pi(\bm{z}_{mis}\mid\bm{z}_{obs}, \bm{y})$, the previous marginal can be
approximated as

$$
\pi(\theta_k \mid \bm{z}_{obs}, \bm{y}) \simeq \frac{1}{L} \sum_{l=1}^L \pi(\theta_k\mid \bm{z}_{obs},  \bm{z}^{(l)}_{mis}, \bm{y}),
$$
\noindent
where $\pi(\theta_k \mid \bm{z}_{obs},  \bm{z}^{(l)}_{mis})$ is the
marginal of $\theta_k$ obtained from fitting the original model with the
observed data and the imputed covariate $\bm{z}^{(l)}_{mis}$.

Note that when continuous covariates with missing values are also present both
approaches can be combined. For example, an imputation sub-model can be
combined for the continuous covariate which is part of the joint model that is
fit to every simulated dataset where only the missing values of the categorical
covariate are filled in. Furthermore, a missingness
sub-model for the categorical variables can be incorporated into the model
similarly to the one used for the continuous variables.

\section{The Integrated Nested Laplace Approximation approach (INLA)}
\label{sec:INLA}

The approach presented in the previous sections overcome a major limitation in INLA, as at present it cannot cope with missing values in covariates. We present here an introduction to the INLA method and the computationa details; then we focus on how to implement our proposed framework. 

INLA \citep{INLAreview:2017, martino:2019, VirBook:2020}  is a deterministic
approach for Bayesian inference. It is designed for the class of latent
Gaussian Markov random field models, where the response $y_{i}$ observed for the $i$-th unit is
assumed to belong to a distribution family (usually part of the exponential
family). This is often characterized by a parameter $\phi_{i}$ defined as a
function of a structured additive predictor $\eta_{i}$ through a link function
such that $g(\phi_{i})=\eta_{i}$ (e.g. the logarithm function is used for
Poisson data). The linear predictor is defined as follows

\begin{equation}
\label{Eq:LinPred}
\eta_i = \beta_0 + \sum_{j=1}^{n_\beta} \beta_j z_{ji} + \sum_{k=1}^{n_f} f^{(k)}(u_{ki}), \qquad i=1,\ldots,n
\end{equation}
where $\beta_0$ is the intercept, the coefficients $\bm
\beta=(\beta_1,\ldots,\beta_{n_\beta})$ quantify the (linear) effect of some covariates
$\bm z= \{\bm{z}_j\}_{j=1}^{n_\beta}$ on the response, and $\bm
f=\left\{f^{(1)}(\cdot),\ldots,f^{(n_f)}(\cdot)\right\}$ is a set of functions
defined in terms of some covariates $\bm u = \{\bm{u}_k\}_{k=1}^{n_f}$. 

Through functions $f(\cdot)$ it is possible to include in the model random
effects (perhaps indexed in space and time), smooth and non-linear effects of
the covariates. For this reason, the class of latent GMRF models can
accommodate a wide range of models, from standard generalized linear models
(GLM) to generalized linear mixed models (GLMM), including data for time
series, lattice data, point pattern and geostatistical data.

As stated, the vector of latent effects $\bm\chi = \{\bm\eta, \beta_0,\bm\beta,\bm
f\}$ is a latent GMRF in the model, which depends on some hyperparameters $\bm
\theta_2$. Moreover, observations are assumed to be independent given the
latent effects $\bm \chi$ and the likelihood hyperparameters denoted by $\bm
\theta_1$. For convenience in the following we will denote the vector of
hyperparameters with $\bm \theta=(\bm \theta_1,\bm \theta_2)$.

The objectives of Bayesian inference with INLA are the marginal posterior distributions
for each element of the latent effects and hyperparameters vector denoted by
$p(\chi \mid \bm y)$ and $p(\theta\mid\bm{y})$, respectively. INLA
provides deterministically accurate approximations to these distributions in a
short computing time by using the Laplace approximation and numerical
integration.


Each latent GMRF model can be rewritten in a hierarchical fashion with three levels:
\begin{enumerate}
\item The model for the observed data  $\bm y=(y_1,\ldots, y_n)$ (i.e. the likelihood) defined as a function of some parameters $\bm \chi$ and hyperparameters $\bm \theta$:
\[
\bm y \mid \bm \chi, \bm \theta \sim \pi(\bm y\mid \bm \chi, \bm \theta)=\prod_{i\in n} \pi(y_i\mid  \chi_i, \bm \theta).
\]


\item The model for the latent effects $\bm \chi$
\[
\bm \chi \mid \bm \theta \sim \text{Normal}\left(\bm 0, \bm Q(\bm \theta)\right)
\]
where $\bm Q(\bm \theta)$ is a sparse precision matrix given the GMRF assumption.

\item The model for the complete vector of hyperparameters: $\pi(\bm \theta)$. As usually hyperparameters are assumed to be independent a priori, $\pi(\bm \theta)$ will be defined as the product of different univariate prior distributions.
\end{enumerate}

Given all these models and components the joint posterior distribution of the random effects and the hyperparameters is given by
\[
\pi(\bm \chi,\bm \theta \mid \bm y)\propto \pi(\bm y\mid \bm \chi,\bm\theta)\pi(\bm \chi \mid \bm \theta)\pi(\bm \theta).
\]

INLA computes the posterior marginals of the hyperparameters and latent effects
using that representation by means of numerical integration and the Laplace
approxiamtion \cite[see][for details]{Rueetal:2009}.

\subsection{Computational details}

The INLA approach is implemented through an \texttt{R} package named 
\texttt{R-INLA},  which is available from the INLA website
(\url{http://www.r-inla.org/home}).  The model to be fit is defined by setting
an formula with all the additive latent effects in the model, which includes
fixed and random effects.  The \texttt{R-INLA} package includes a good number of
implemented latent effects but others can be implemented as well \citep[see,
for example][]{VirBook:2020}.  Note that by default, when \texttt{R-INLA} finds
missing values in the covariates (which have the value \texttt{NA} in
\texttt{R}) they are replaced by zeros so that the effect of the covariate does
not affect the linear prediction of that subject. However,
this is an issue that could result in biased estimates of the coefficients
of the covariates. This is described in the
\texttt{R-INLA} list of frequently asked questions (FAQ) in the package
website.  If the missing value is found in the response variable, the
predictive distribution is computed.

Generic latent effects can be implemented by defining their structure as a
latent GMRF. This means definining the mean, precision, hyperparameters and the
priors of the hyperparameters. These are known as \texttt{rgeneric} latent
effects in \texttt{R-INLA} \citep[see, for example][Chapter 11]{VirBook:2020}. Once a
new latent effect is defined, it can be easily incorporated as any other
additive effect in the model formula.

For the new latent effects described in this paper, note that it is
defined as in equation~\eqref{Eq:newlatenteff}, so that the only
difference will be in how the mean $\bm \mu_{c}$ and precision $\bm Q_c$
of the block of the missing values is defined. Remember that the block
of the observed covariates is simply there to make those values of
the latent effect to be as close as possible to the observed values
and that it does not depend on any hyperparameter or other data.

Furthermore, the role of the prior on the hyperparameters of the imputation
model $\bm \theta_I$ is now taken by distribution $\pi(\bm\theta_I \mid
\bm{z}_{obs})$. Hence, the actual prior used in the latent effects is taken as

$$
\pi(\bm\theta_I \mid \bm{z}_{obs}) \propto \pi(\bm z_{obs}\mid \bm\theta_I)\pi(\bm \theta_I)
$$
\noindent
and the normalizing constant is ignored as it is not needed. In a typical
implementation of a latent effect, the prior of $\bm\theta_I$ would be a
typical distribution density that depends on a set of fixed hyperparameters,
but now the prior of $\bm\theta_I$ is made of the product of the two terms
above.  For this reason, it can be regarded as an informative prior as it is
essentially estimated from a model fit to $z_{obs}$. This is what will allow
the latent effect to produce good estimates of the missing values (if the
imputation model is correct).  In general, there is no way to assess this, but
the more covariates used in the imputation model the better \citep[see][Chapter
25]{GelmanHill:2007}. 

The actual prior of the model hyperparameters is $\pi(\bm \theta_I)$ and this
can take different forms depending on the number and type of hyperameters
in the model. Usually, this will be split into the product of several univariate
prior distributions.

Note also that \texttt{R-INLA} works with unbounded hyperparameters, so that
the parameters in $\bm \theta_I$ may need to be transformed when the latent
effect is defined. This may also require to include additional terms
in the prior \citep[see, for example][Chapter 11]{VirBook:2020}.
A typical example is to use internally the log-precision instead of the
precision.

Once the imputation latent effect is included in the model formula
it will be part of the joint latent effect $\bm \chi$ and incorporated
into the Bayesian model, so that a full Bayesian approach is used to 
estimate all the model parameters.

As stated in previous sections, a missingness sub-model can be
included (in addition to an imputation one) for the case in which
missingness is MAR or MNAR. Including a missingness model requires defining a model with two likelihoods:
one for the main model and a binomial model for the missingness indicator
variables. Note that under MCAR and MAR both models are independent, hence the latter is not needed; however, under MNAR it is necessary to explicitly include it and to make it dependent on the variables with imputed values.
Hence, there will be feedback between both models that may affect the imputation
process and the estimation of the other model parameters.

Full details about how to fit these models in \texttt{R} are provided in the
Supplementary materials together with the associated \texttt{R} code for the examples
developed in Section~\ref{sec:examples}.

\section{Examples}
\label{sec:examples}

In this section we develop two examples to show how the imputation method
proposed above can be used with INLA under MCAR, MAR and MNAR.  The first
example shows  typical regression model in biostatistics with real missing
data. This is useful to show how a typical multiple linear regression can be
used for multiple imputation.  The second one is based on spatially correlated
data to assess the performance of our proposal on a simulated study in which a
spatially correlated covariate is missing.  Note that the aim is not to provide
a comprehensive analysis of the dataset with missing values but to illustrate
the methods described in this paper.

All models have been fit with INLA and its associated R package
\texttt{R-INLA}.  The latent effects required to impute the missing values of
the covariates are implemented in  a new \texttt{R} package called
\texttt{MIINLA} which is available from the Github repository at
\mbox{\url{https://github.com/becarioprecario/MIINLA}}.
The \texttt{R} code to run the examples is available from
\mbox{\url{https://github.com/becarioprecario/MIINLA_paper}}.

\subsection{Imputation using linear models}

The \texttt{nhanes2} dataset \citep{Schafer:1997} in the mice \texttt{R}
package \citep{micepkg} records data on 25 participants in the National Health
and Nutrition Examination Survey (NHANES). Variables in the dataset include
body mass index, cholesterol level, age group and hypertensive status. The
dataset presents missing observations in body mass index, hypertensive status
and cholesterol level.

We will use this dataset to build a model to explain cholesterol level on age
group and body mass index, where this is imputed. The imputation model will be
based on a linear regresion on the age group. There are three age groups 20-39,
40-59 and 60+ years, and the first group will be set as the reference level.

It is worth noting that having missing values in the response variable (i.e.,
cholesterol level)  is not a problem as the predictive distribution can be
easily computed with INLA. Hence, the output from fitting this model will
include the posterior distribution of the imputed values as well as
the predictive distribution for the missing responses.

The analysis model is the following:

$$
chol_i = \alpha + \beta_1 age^{40-59}_i + \beta_2 age^{60+}_i + \beta_3 bmi_i + \varepsilon_i,\ i=1,\ldots,25
$$
\noindent
where $chol_i$ refers to the cholesterol level, $bmi_i$ to the body mass index,
$age^{40-59}_i$ and $age^{60+}_i$ are indicator variables of age for groups
40-59 and 60+, respectively, and $\varepsilon_i$ is a Gaussian error term with
zero mean and precision $\tau$.

Note that the missing values of $bmi_i$ are obtained from the imputation model 
based on linear regression discussed above using as predictors variables
$age^{40-59}_i$ and $age^{60+}_i$. The imputation model is specified as

$$
bmi_i = \alpha_I + \beta_{I1} age^{40-59}_i + \beta_{I2} age^{60+}_i +\varepsilon_{Ii},\ i\in\mathcal{I}.
$$
\noindent
Here, $\mathcal{I}$ represents the set of indices of the observations with
missing values of body mass index. Parameters $\alpha_I$, $\beta_{I1}$,
$\beta_{I2}$ represent the intercept and the covariate coefficients used in the imputation model, and $\varepsilon_{Ii}$ is a
Gaussian error with zero mean and precision $\tau_I$.  Note that all the
parameters in the imputation model are mainly informed from the observed values
of the body mass index and age, and their prior distributions. Because the
imputation model is part of the joint model there is also feedback from all the
other parts of the model when estimating the imputation model parameters and
the imputed values of body mass index.

Finally, a logistic regression is used on the missingness status of $bmi_i$
using a logistic regression with a particular linear predictor. We tried
three different approaches to assess missingness under MCAR, MAR and MNAR.
Under MCAR, the linear predictor has simply an intercept term, under MAR it is
an intercept plus the covariate of age group, and under MNAR it is the
intercept plus covariate $bmi_i$ (that includes the imputed values). 
The coefficient of $bmi_i$ would indicate whether the
values of $bmi_i$ have been missed completely at random or following a different
scheme.

This model can be represented as

\begin{eqnarray}
M_i & \sim & Bernoulli(p_i),\ i = 1\ldots,25\nonumber\\
logit(p_i) & = & \alpha_M + \beta_{M1} age^{40-59}_i + \beta_{M2} age^{60+}_i + \delta bmi_i
\end{eqnarray}
where $M_i$ is a missingness indicator for $bmi_i$ (0 for observed and 1 for
missing).  The priors for the coefficients of the fixed effects are independent
Normal distributions with zero mean and precision 0.001.  For the precision
parameters, a Gamma with parameters 1 and 0.00005 is used to provide a vague
prior. All parameters are considered to be independent a priori.

\begin{table}
\caption{\label{tab:nhanesmodels}Posterior mean (and standard deviation) of the parameters from the joint models in the nhanes2 dataset.}
\centering
\small
%
\begin{tabular}{c|c|r|r|r}
 & & \multicolumn{3}{c}{Missingnesss mechanism in the model}\\
\cline{3-5}
Sub-Model & Parameter & MCAR & MAR & MNAR \\ 
\hline
\multirow{5}{*}{Analysis} & 
   $\alpha$ & -4.084 (1.209) & -4.233 (0.816) & -4.864 (1.247)\\
 & $\beta_1$ & 1.145 (0.421) & 1.154 (0.398) & 1.229 (0.447)\\
 & $\beta_2$ & 1.866 (0.541) & 1.879 (0.501) & 1.940 (0.580) \\
 & $\beta_3$ & 0.111 (0.049) & 0.145 (0.044) & 0.156 (0.044)\\
 & $\tau$ & 2.219 (0.786) &  2.568 (1.312)   & 2.620 (1.169)\\
\hline
\multirow{4}{*}{Imputation} & 
   $\alpha_I$ & 31.195 (1.569) & 30.046 (1.515) & 30.401 (1.296)\\
 & $\beta_{I1}$ & -5.902 (1.985) &  -5.204 (2.316) & -4.711 (1.742)\\
 & $\beta_{I2}$ & -7.395 (1.733) & -5.561 (2.372) & -6.153 (2.126)\\
 & $\tau_I$ & 0.058 (0.027) & 0.073 (0.023) & 0.096 (0.030) \\
\hline
\multirow{3}{*}{Missingness} & 
  $\alpha_M$ & -- & -0.337 (0.585) & -4.633 (4.892)\\
 & $\beta_{M1}$ & -- & 1.879 (0.501) & --\\
 & $\beta_{M2}$ & -- & -0.377 (1.044) & --\\
 & $\delta$ & -- & -- & 0.092 (0.167)\\
\end{tabular}

\end{table}

Table~\ref{tab:nhanesmodels}  shows the different estimates for all the models
considered. Regarding the main Gaussian sub-model, it seems that all three
covariates included in the model play a significant role when explaining
cholesterol level. In addition, point estimates are very similar across
different missingness mechanisms.  In the imputation sub-model, we also observe
that point estimates are very similar across missingness mechanisms. Age also
plays an important role when imputing the missing values of body mass index.
Finally, the different sub-models for the missingness mechanism are not 
directly comparable. Under MCAR, parameter $\alpha_M$ estimates the log-proportion of missing values in the covariate. Under MAR, $age^{40-59}$ helps
to explain why some values of body mass index are missing. Lastly, under
MNAR the missing values do not appear to depend on their actual values
as the estimate of $\delta$ is close to zero.

Cholesterol level seems to
increase with age. In addition, the imputation models points to that body mass
index seems to decrease with age.  Although this is counterintuitive, we
believe that is due to the general pattern observed in the dataset, which
contains data on 25 people and only 13 of them have completely observed
covariates.

As a final remark, it is worth noting that fitting these models took a few
seconds. Hence, the sensitivity analysis could include other models than the
ones presented here. See, for example, \citet{mason_strategy_2012} for a
general discussion and alternative models for the sensitivity analysis.
Larger datasets may take longer to run, but INLA will be able to fit
these models faster than typical MCMC algorithms.

\subsubsection{Imputation of categorical covariates with missing values}

As we have mentioned in the description, this dataset includes an
indicator of hypertensive status of the subjects. This categorical covariate
also contains several missing values. To illustrate how missing values in
continuous and categorical covariates can be handled at the same time we 
fit a model in which body mass index and hypertensive status are included. The
imputation of body mass index will be done within the joint model as previously described, but the imputation of hypertension will be done
using a multiple imputation approach; this means that an imputation model will be fit for
hypertension, values of hypertensive status sampled from this model and used to fill
the gaps in the original dataset. This will provide a number of complete
datasets to which the analysis model will be fit; then the results will be
pooled to obtain final estimates using Bayesian
model averaging with equal weights \citep{GomezRubioetal:2019}.

The analysis model becomes:

$$
chol_i = \alpha + \beta_1 age^{40-59}_i + \beta_2 age^{60+}_i + \beta_3 bmi_i + \beta_4 hyp_i + \varepsilon_i,\ i=1,\ldots,25
$$

For simplicity, the missingness mechanism will not be asessed now. This
implies assuming MCAR, but we have already seen that the model estimates
will be close to model fit under MAR and MNAR for the case of body mass
index.

\begin{table}
\caption{\label{tab:hyp} Posterior probabilities of being hypertensive for the different age groups.}
\begin{tabular}{c|c|c|c}
 & \multicolumn{3}{c}{Age group} \\
Hypertensive & 20-39 & 40-59 & 60+\\
\hline
Yes & 1.00 & 0.66 & 0.49\\
No  & 0.00 & 0.34 & 0.51\\
\end{tabular}
\end{table}

The imputation model for hypertensive status ($hyp_i$) will be a 
multinomial model fit using the multinomial-Poisson transformation
\citep{Baker:1994}. This will provide estimates of the 
posterior probabilities of being hypertensive given the age group,
which will be used to impute the missing values according to the
age group of the patient.  These posterior probabilities are
shown in Table~\ref{tab:hyp}. Note that in this particular case
a logistic regression would have been enough, but we have preferred
to use the multinomial-Poisson transformation because it is a more
general approach for the case of more than two categories.

We have drawn 100 samples to fill in the missing values of the hypertensive
status, so that 100 different completed datasets have been used to fit
the model. The resulting models have been pooled to obtained the posterior
marginals of the model parameters using Bayesian
model averaging with equal weights \citep{GomezRubioetal:2019}. These are shown in Table~\ref{tab:hypmodel}.

\begin{table}
\caption{\label{tab:hypmodel} Estimates of the model parameters using multiple imputation on body mass index and hypertensive status.}
\begin{tabular}{c|c}
\multicolumn{2}{c}{Gaussian model}\\
\hline
Parameter & Estimate \\
\hline
 $\alpha$ & -4.981 (1.166)\\
 $\beta_1$ & 1.208 (0.518) \\
 $\beta_2$ &1.985 (0.635) \\
 $\beta_3$ & 0.134 (0.072) \\
 $\beta_4$ & 0.027 (0.566)\\
 $\tau$ & 1.965 (0.994) \\
\hline
\hline
\multicolumn{2}{c}{Imputation model for $bmi_i$} \\
\hline
$\alpha_I$ & 29.612 (1.474)\\
$\beta_{1I}$ & -3.899 (2.114)\\
$\beta_{2I}$ & -6.116 (2.337)\\
$\tau_I$ & 0.092 (0.034)\\
\end{tabular}
\end{table}

As expected, the estimates of the coefficients of age are close to the ones in
the previous models. The coefficient of hypertensive status is close to zero,
which indicates no association between cholesterol level and hypertensive
status. Furthermore, the imputation model for body mass index based on a linear
regression on age provides similar estimates to the imputation models fit
previously and with similar effects of age on body mass index.

\subsection{Simulation study: Imputation of correlated data}

The second example that we present is a simulation study based on the North
Carolina Sudden Infant Death Syndrome (SIDS) dataset. It records several data,
which includes the number of sudden infant deaths per county in the period 1974-78
($O_i$), the total number of births ($N_i$), as well as the number of non-white
births ($NW_i$).  The expected number of cases in each county ($E_i$) can be
obtained using internal standardization, so that the standardized mortality
ratio (SMR) can be computed as $O_i / E_i$. Furthermore, several authors
\citep[see, for example,][]{Cressie:2015} have described the strong spatial
pattern in the data, in the relative risk (estimated using the SMR, for
example) and its correlation with the proportion of non-white births.

The model of interest to be fit is simply a Poisson regression, as follows:

$$
O_i \sim Po(\mu_i); \mu_i = E_i \theta_i,\ i=1,\ldots, 100
$$

$$
\log(\theta_i) = \alpha + \beta \; nwpropbirths_i
$$

Here, the covariate $nwpropbirths_i$ is the logit of the proportion of
non-white births ($NW_i$), that has been re-centered and re-scaled so that it
is not bounded. This derived covariate has still a strong spatial pattern
and a high correlation with the SMR.

\begin{figure}
\centering
\includegraphics[width=7cm]{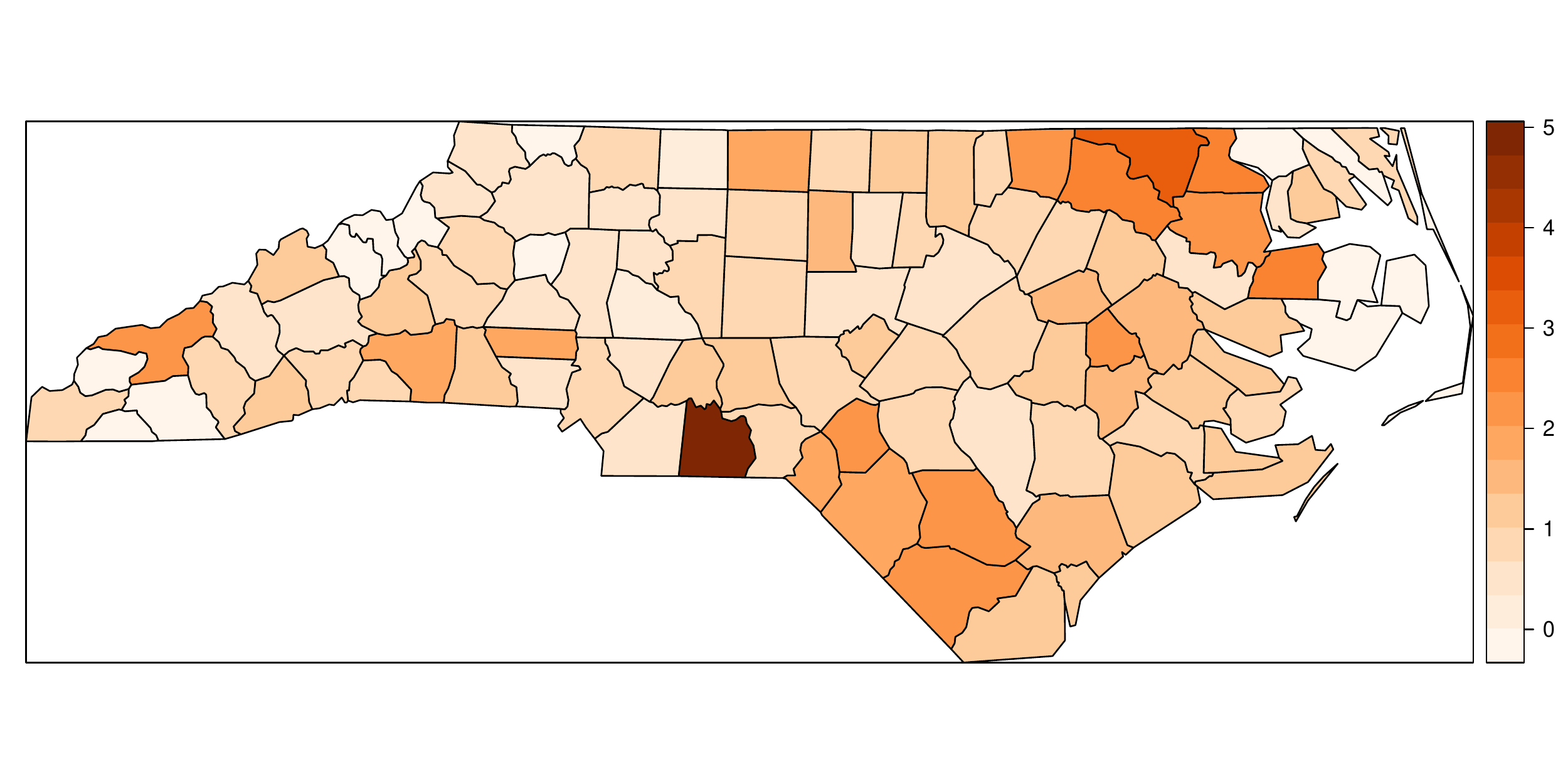}
\includegraphics[width=7cm]{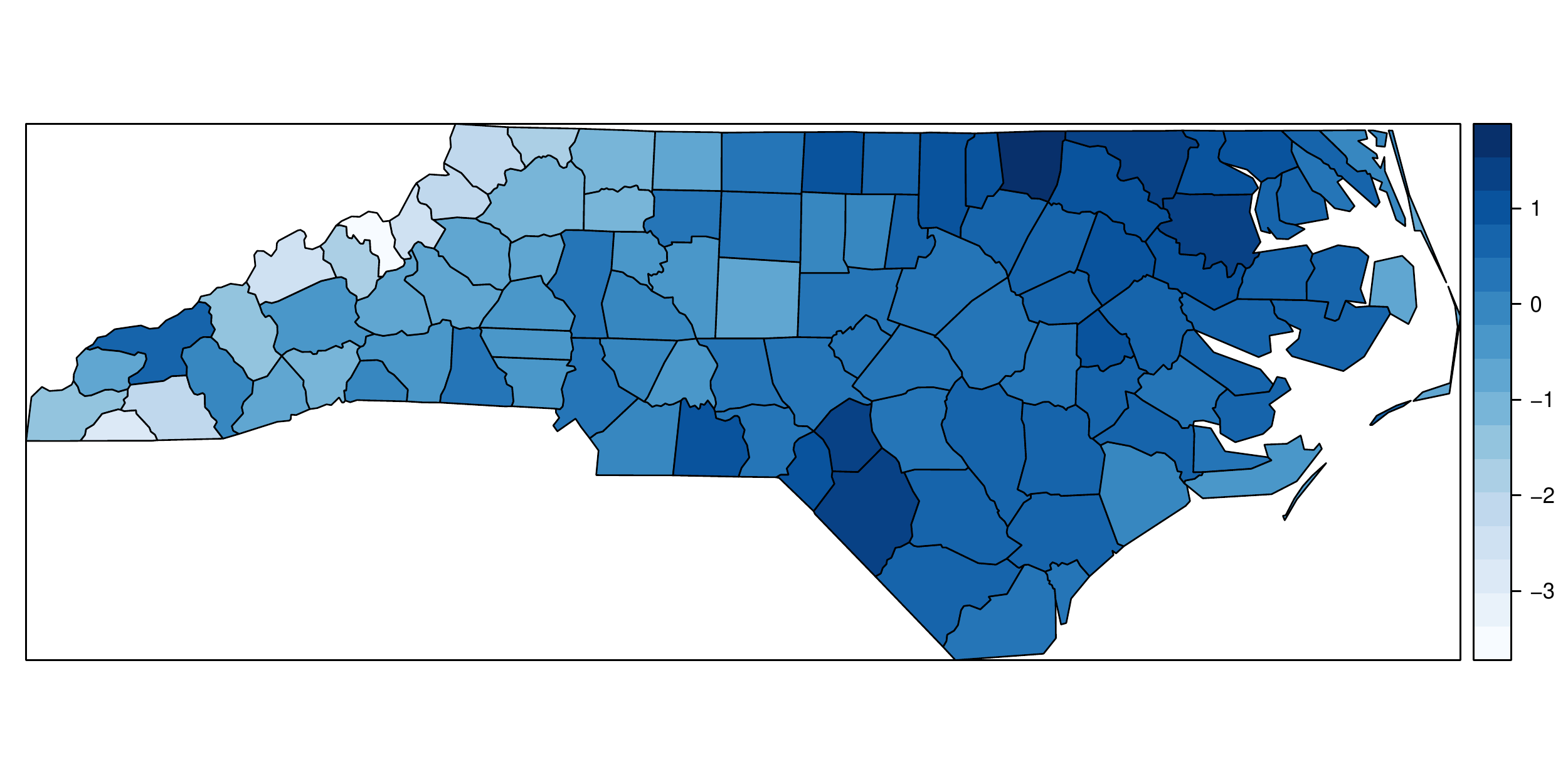}
\caption{Standardized mortality ratio (SMR, left) and proportion of non-white births (right) in North Carolina in the period 1974-78.}
\label{fig:NCSIDS}
\end{figure}

Figure \ref{fig:NCSIDS} shows the SMR for the period 1974-78 and the
transformed proportion of non-white births ($nwpropbirths_i$). The SMR shows some
areas of high risk and a strong correlation with the proportion of non-white
births. Hence, this covariate can be useful when building models
to explain the spatial variation of SIDS in North Carolina.

The simulation study will remove 5\%, 10\%, 15\%, 30\%  and 50\% of the
covariate values (i.e., proportion of non-white births) using MCAR and MNAR
mechanisms. Note that MAR can be regarded as an extension to MCAR that
considers other observed covariates in the linear predictor of the logistic
regression in the imputation model. Although MAR may seem more
reasonable, it is simply a matter of including other covariates in the linear
predictor of the missingness model so it is computationally feasible but
it adds little to the comparisson. This is why we have not considered it.

The missing observations will be nested accross the five scenarios, i.e., the
observations removed in the 10\% secenario will also be removed in the 15\%
scenario and so on.  Furthermore, the probability of being missing under the
MNAR mechanism $p_i$ is

$$
logit(p_i) = \alpha_M + 5 x_i
$$
\noindent
where $\alpha_M$ is set as the logit of 0.5 and $x_i$ represents the value
of the covariate with missing values.

This simulation is intended to compare mild to severe missingness under five
different scenarios for MCAR and MNAR. Models will be fit assuming MCAR and
MNAR missingness, so that we fit 20 models in total. Under MCAR, we only
fit the analysis and imputation model. Under MNAR, in addition we will
assess whether the joint approach including the missingness mechanism is able to capture the type of missingness.

\begin{figure}
\centering
\includegraphics[width=14cm]{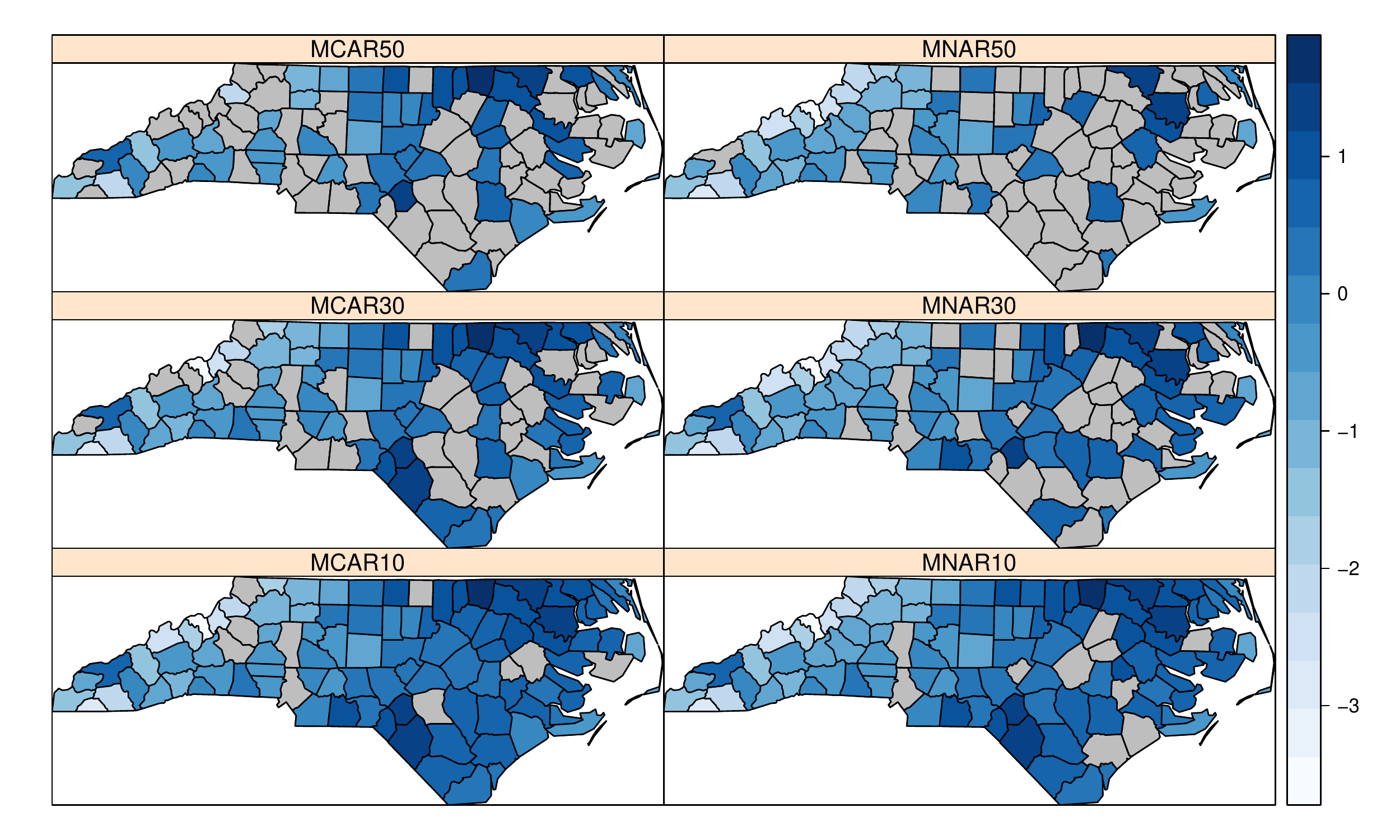}
\caption{Missing observations (in grey) of the proportion of non-white births.}
\label{fig:NCSIDSmissing}
\end{figure}

Figure \ref{fig:NCSIDSmissing} shows the missing values of the proportion of
non-white births for three of the scenarios considered in this simulation
study. As it can be seen, when the percentage of missing values is 50\% under
MNAR missing values concentrate in the counties with high values of the
covariate.

In addition, the imputation model proposed is based on the conditional autoregressive specification presented in Section~\ref{Sec:imputationCAR}, so that imputation is included within the main model.  This imputation
model will have the following parameters: $\tau_I$ is the precision of the CAR
specification, $\rho_I$ the spatial autocorrelation and $\alpha_I$ the mean value of
the covariates.

Finally, a logistic regression on the missingness variable $m_i$ (0 for
observed and 1 for missing) is used to model the missingness mechanism
(under MNAR):

\begin{eqnarray}
m_i & \sim &  Bernoulli(p_i);\ i=1,\ldots, 100 \nonumber\\
logit(p_i) & = & \alpha_M + \beta_M  nwpropbirths_i \nonumber\\
\end{eqnarray}

Note that the imputed values appear both in the Poisson regression and the
sub-model on the missingness mechanism. Non-zero values of $\beta_M$ indicate
that the probability of being missing depends on the actual values.

Table \ref{tab:NCSIDSMCAR}  summarizes the models fit to the data under MCAR.
Here, an imputation sub-model for body mass index has been included but not a
joint model for the missingness as under MCAR it is not necessary.  In general,
there are not large differences between the different models fit to the
datasets regarding percentage of missing values and type of actual missingness.
However, these differences become larger as the proportion of missing values
increases, which was to be expected. These differences are noticeable for the
case of 50\% of missing values both under MCAR and MNAR.

The estimates of the imputation models are quite similar as well,
across missingness type in the data and proportion of missing values.
However, some differences are observed for 30\% and 50\% of missing values.
In particular, the estimates of $\alpha_I$ differ.

Table~\ref{tab:NCSIDSMNAR} summarize the (joint) models fit to the data
considering a MNAR scenario. This includes the model fit to the complete
dataset, and the binomial sub-model in the joint model to assess the
missingness mechanism.  First of all, the posterior distribution of $\beta_M$
helps to determine the missingness mechanism. Its posterior estimate is very
close to zero under MCAR, while it is above zero under MNAR (but for the case
of 5\% of missing values). It is worth stating that we can assess this now
because this is simulated data and the true missingness mechanism is known.

Regarding the imputation model, the estimates are very similar across
scenarios. Finally, the estimates of the parameters in the Poisson model
are in general very close to the model fit to the full dataset.

It is worth noting that under MNAR with 50\% of missing observations the point
estimates of the parameters in the Poisson sub-model show the largest departure
from the model fit to the full dataset. This is probably due to the fact that
the imputation model is not able to fully recover the values of the covariates
as missing values tend to have high values and there is not enough information
in the observed values as to recover this pattern.

To sum up, we believe that the imputation model behaves as expected and
provides a good performance in all cases. Most importatly, the joint model is
able to identify between MCAR and MNAR situations as well as imputing the
covariates and fit the model of interest to the data. Again, ths is possible
now because the missingness mechanism is known but in real applications
we would propose different models and conduct a sensitivity analysis.

When the models fit under MCAR (Table \ref{tab:NCSIDSMCAR}) and under MNAR
(Table~\ref{tab:NCSIDSMNAR}) are compared, it should be mentioned that when
data under MCAR are analysed both models produce very similar results because
the missingness mechanism is independent of the observed data.  For the
analysis of the data simulated under MNAR, differences can be observed because
now the missingness mechanism depends on the covariate (including the imputed
data) and the estimates of the parameters in the imputation sub-model are
different.

\begin{table}
\caption{\label{tab:NCSIDSMCAR}Posterior mean (and standard deviation) of the parameters of the imputation models fit to the data under MCAR.}
\centering
\tiny


\begin{tabular}{c|c|c|c|c|c|c|c|c}
 & & \multicolumn{6}{c}{Model under MCAR}\\
\cline{3-9}
 & & \multicolumn{2}{c|}{Poisson} & \multicolumn{3}{c|}{Imputation}& \multicolumn{2}{c}{Missingness}\\
\hline
Missingness & \% missing & $\alpha$ & $\beta$ & $\tau_I$ & $\rho_I$ & $\alpha_I$ & $\alpha_M$ & $\beta_M$\\
\hline
-- & 0 & -0.141 (0.046) & 0.524 (0.068) & -- & -- & -- & -- & --  \\
\hline
MCAR & 5 & -0.126 (0.047) & 0.518 (0.068) & 2.129 (0.305) & 0.977 (0.022) & -0.211 (0.162) &  -- & -- \\
MCAR & 10 & -0.114 (0.047) & 0.496 (0.069) & 2.076 (0.301) & 0.976 (0.024) & -0.215 (0.165) & -- & --\\
MCAR & 15 & -0.120 (0.048) & 0.504 (0.067) & 1.915 (0.294) & 0.973 (0.027) & -0.234 (0.175) & -- & --\\
MCAR & 30 &  -0.099 (0.049) & 0.507 (0.065) & 1.776 (0.295) & 0.960 (0.039) & -0.175 (0.183) & -- & --\\
MCAR & 50 & -0.077 (0.051) & 0.518 (0.070) & 2.461 (0.481) & 0.957 (0.044) & 0.034 (0.169) & -- & --\\
\hline
MNAR & 5 & -0.131 (0.045) & 0.506 (0.067) & 2.040 (0.292) & 0.977 (0.022) & -0.236 (0.166) & -- & --\\
MNAR & 10 & -0.138 (0.048) & 0.506 (0.068) & 1.991 (0.288) & 0.976 (0.023) & -0.220 (0.167) & -- & --\\
MNAR & 15 & -0.110 (0.048) & 0.495 (0.068) & 1.966 (0.289) &  0.976 (0.024) & -0.238 (0.170) & -- & --\\
MNAR & 30 & -0.105 (0.050) & 0.453 (0.070) & 1.827 (0.291) & 0.975 (0.025) & -0.342 (0.189) & -- & --\\
MNAR & 50 & -0.064 (0.055) & 0.419 (0.061) & 1.421 (0.279) & 0.964 (0.037) & -0.423 (0.226) & -- & --\\
\hline
\end{tabular}
\end{table}

\begin{table}
\caption{\label{tab:NCSIDSMNAR}Posterior mean (and standard deviation) of the parameters of the imputation models fit to the data under MNAR.}
\centering
\tiny


\begin{tabular}{c|c|c|c|c|c|c|c|c}
 & & \multicolumn{6}{c}{Model under MNAR}\\
\cline{3-9}
 & & \multicolumn{2}{c|}{Poisson} & \multicolumn{3}{c|}{Imputation}& \multicolumn{2}{c}{Missingness}\\
\hline
Missingness & \% missing & $\alpha$ & $\beta$ & $\tau_I$ & $\rho_I$ & $\alpha_I$ & $\alpha_M$ & $\beta_M$\\
\hline
-- & 0 & -0.141 (0.046) & 0.524 (0.068) & -- & -- & -- & -- & --  \\
\hline
MCAR & 5 & -0.121 (0.047) & 0.512 (0.068) & 2.120 (0.305) & 0.977 (0.022) & -0.217 (0.163) & -3.218 (0.565) & -0.514 (0.465) \\
MCAR & 10 & -0.111 (0.048) & 0.494 (0.069) & 2.073 (0.301) & 0.975 (0.024) & -0.216 (0.165) & -2.271 (0.349) & -0.074 (0.392)\\
MCAR & 15 & -0.127 (0.049) & 0.505 (0.067) & 1.903 (0.293) & 0.972 (0.027) & -0.218 (0.176) & -1.821 (0.309) & 0.359 (0.396)\\
MCAR & 30 & -0.110 (0.050) & 0.507 (0.065) & 1.768 (0.294) & 0.960 (0.039) & -0.141 (0.187) & -0.896 (0.232) & 0.339 (0.309) \\
MCAR & 50 & -0.079 (0.054) & 0.518 (0.070) & 2.458 (0.480) & 0.956 (0.044) & 0.040 (0.176) & -0.014 (0.203) & 0.038 (0.307) \\
\hline
MNAR & 5 & -0.132 (0.045) & 0.502 (0.068) & 2.046 (0.293) & 0.976 (0.023) & -0.236 (0.165) & -3.286 (0.720) & 0.810 (0.795) \\
MNAR & 10 & -0.153 (0.049) & 0.486 (0.071) & 1.964 (0.287) & 0.977 (0.022) & -0.225 (0.170) & -2.947 (0.849) & 1.661 (0.828)\\
MNAR & 15 & -0.133 (0.049) & 0.481 (0.069) & 1.928 (0.287) & 0.977 (0.023) & -0.227 (0.173) & -2.225 (0.529) & 1.306 (0.592)\\
MNAR & 30 & -0.152 (0.052) & 0.423 (0.069) & 1.688 (0.285) & 0.976 (0.024) & -0.190 (0.200) & -1.385 (0.450) & 1.477 (0.492) \\
MNAR & 50 & -0.172 (0.060) & 0.380 (0.060) & 1.230 (0.266) & 0.969 (0.032) & -0.093 (0.253) & -0.303 (0.351) & 1.576 (0.434)\\
\hline
\end{tabular}
\end{table}

\begin{figure}
\caption{\label{fig:imputed} Posterior marginal distributions of some of the imputed values for missingness of 50\% under MNAR. The lines represent the actual value (solid vertical line),  the posterior marginal from the MCAR model (dashed line) and the posterior marginal from the MNAR model (dotted line). The value between parenthesis corresponds to the proportion of missing values in the neighbour counties.}
\centering
\includegraphics[width=14cm]{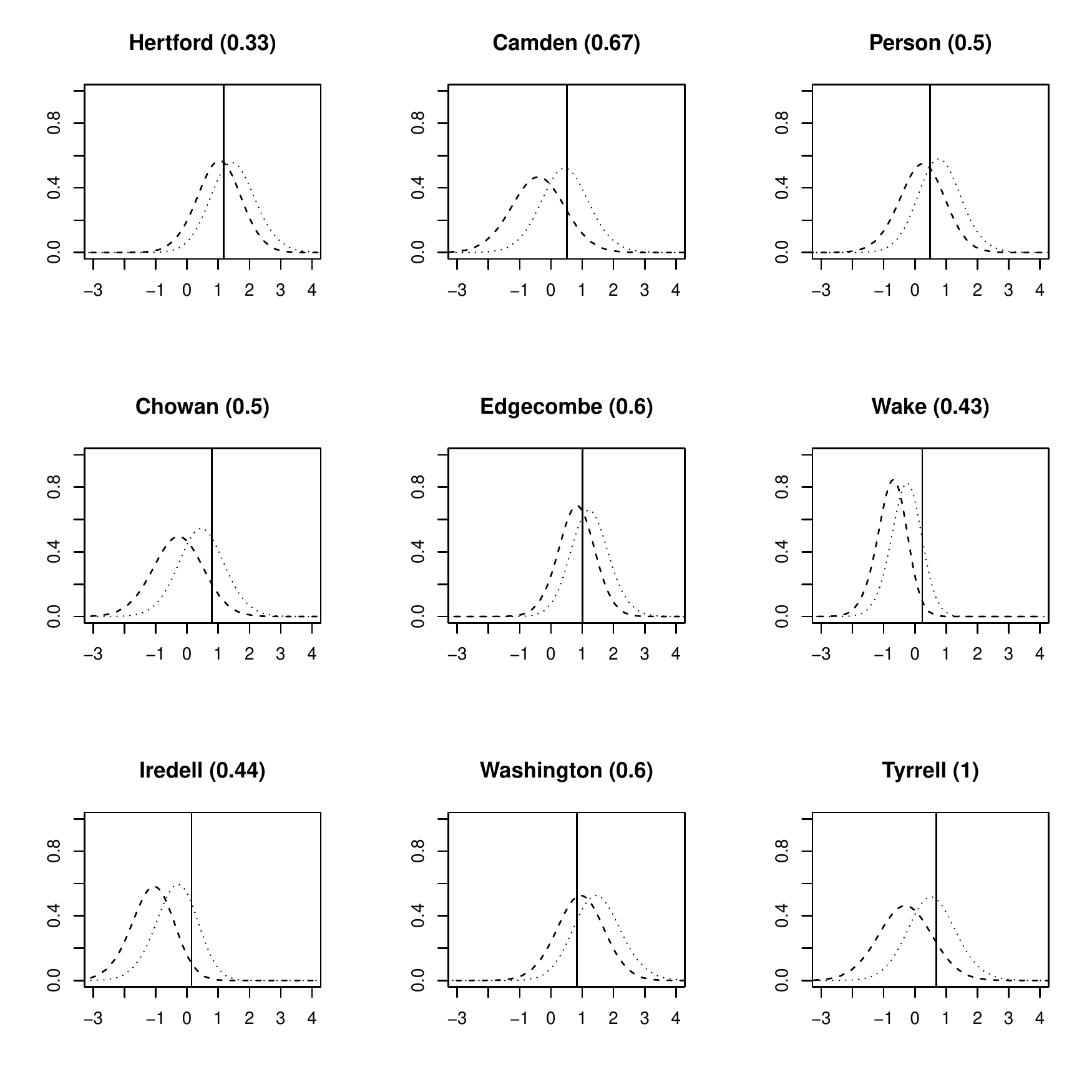}
\end{figure}

Finally, we have included the posterior distributions of some imputed values of
the covariate in Figure~\ref{fig:imputed}. In particular, we have considered
the dataset with 50\% missing values under MNAR and taken nine counties with
missing values that have missing values also in the simulated data under MCAR.
This produces a set of counties with a wide variety in the posterior marginals
of the imputed values.  The posterior marginals shown are for the imputation
model under MCAR in Table~\ref{tab:NCSIDSMCAR} (dashed line) and the imputation
model under MNAR in Table~\ref{tab:NCSIDSMNAR} (dotted line).  The vertical
solid line shows the actual value of the missing covariate.  Furthermore, we
have kept the same axes scale in all plots so that differences are appreciated
better.

In general, both marginals are close in all cases. Under MNAR (dotted
lines), the posterior mode seems to be closer to the actual value 
for most of the counties in the plot. This should not be surprising as
this is the actual missingness mechanism in the data.

It is worth noting that when the missing data obtained under MCAR are considered
the posterior marginals of the analogous models look the same in the plots.
This shows that handling imputation of missing values with INLA is an
interesting way to conduct sensitivity analysis.

\section{Discussion}
\label{sec:discussion}

We have shown how the general problem of dealing with missing observations
in the covariates and performing multiple imputation under different missingness
mechanisms can be recast within the framework on latent Gaussian Markov
random field models. This has the main advantadge that models expressed
as latent GMRFs can be fit with the INLA methodology for speed. Furthermore,
this fills an important gap in the INLA methodology as now models with
missing values in the covariates can be easily fit.

Imputation models for the covariates can also take many different forms when
defined as GMRFs. In this work we have only considered a linear regression model
and spatially correlated model for imputation, but other similar imputation
models could be easily developed.  For example, these could tackle
missing observations in longitudinal data or time series.
Furthermore, the methods proposed can be extended to consider imputation
of more than one covariate at the same time by relying on multivariate
Gaussian models.

The implementation of the multiple imputation models take the form of new
latent effects for the \texttt{R-INLA} package and they are available within
the \texttt{MIINLA} package for the \texttt{R} programming language. These new
latent effects have been developed using the \texttt{rgeneric} framework for
latent effects development within the \texttt{R-INLA} package.  Nonetheless,
this approach could be implemented in any other software packages for Bayesian
inference.

Although we have focused on imputation of continuous covariates, missing values
in categorical covariates can also be handled. However, as stated in the paper,
this case does not fit within the paradigm of latent GMRF models easily.
However, INLA can be used to propose an imputation model for the missing
categorical data and to fit the model of interest to these full datasets. The
fitted models can then be combined to account for the uncertainty of the imputed
values in the estimation of the model parameters using Bayesian model
averaging.

When the missing values of the categorical covariates index a latent effect 
the imputation of missing values becomes more complex. This is the
case, for example, when random effects are estimated for different groups
in the data using multilevel models. However, this scenario could also be
handled using the multiple imputation methods described in this paper.

In addition to handling and imputing missing values, this new framework allows
us to consider the missingness mechanism using a joint model fit within the
INLA methodology. Hence, the analysis of data with missing observations can now
be completely carried out within the INLA framework. 

Sensitivity analysis on the missingness mechanism, required when it is not
ignorable, can benefit from the the computational speed of the INLA method.
First of all, models are fit faster than with typical MCMC methods, which helps
to define the scenarios to test. Secondly, more scenarios can be tested as the
time required to fit the models is reduced.

%
%
%

\section*{Acknowledgements}

V. G\'omez-Rubio has been 
supported by grant MTM2016-77501-P from the Spanish Ministry of Economy and Competitiveness co-financed with FEDER funds, and grant SBPLY/17/180501/000491 funded by Consejer\'ia de Educaci\'on, Cultura y Deportes (JCCM, Spain) and FEDER. Marta Blangiardo acknowledges partial support through the grant R01HD092580 funded by the National Institute of Health.


\bibliographystyle{rss}
\bibliography{missingdata}

\end{document}